\documentclass[12pt]{article}

\usepackage{amsmath, setspace}
\usepackage{amssymb, xspace, mathtools}

\usepackage{graphicx}
\usepackage{color} 

\usepackage{subcaption}
\usepackage{hyperref}
\usepackage{times}
\usepackage[backend=biber,style=nature]{biblatex}
\usepackage[table,rgb]{xcolor}
\usepackage{multirow}
\usepackage[misc,geometry]{ifsym}

\usepackage{caption}
\DeclareCaptionLabelSeparator{pipe}{$\; | \;$}
\definecolor{abstarctBlue}{rgb}{0.0706, 0.349, 0.6667} 
\definecolor{abstarctBlue2}{cmyk}{ 0.5143,   0.2857,    0.0286,    0.3714}
\definecolor{Section}{cmyk}{0.2,0.8,0.8,0.3}

\usepackage{soul}
\sodef\an{\fontfamily{phv}\selectfont}{.08em}{1em plus1em}{0.5em plus.1em minus.1em} 
\sodef\ann{\fontfamily{phv}\selectfont}{0.04em}{0.5em plus0.02em}{0.1em plus.1em minus.1em}

\makeatletter
\renewcommand{\@biblabel}[1]{\quad#1.}
\makeatother

\date{}

\usepackage{overpic}
\newcommand*{\hvfont}{\fontfamily{phv}\selectfont}

\newcommand{\inb}[2]{ \begin{overpic}[width = .42\textwidth]{#1} \put(0,75){\large \bf \hvfont #2}\end{overpic} }

\usepackage{soul}
\sodef\an{\fontsize{13}{14}\fontfamily{phv}\selectfont}{.10em}{1em plus1em}{2em plus.1em minus.1em}

\definecolor{shadecolor}{RGB}{240,240,250}
\newcommand{\graybox}[1]{\colorbox{shadecolor}{#1}}

\newcommand{\supp}{Supplemental Information\xspace	} 

\newcommand{\fg}{\textcolor{linkcolor}{Fig.} \ref}
\newcommand{\figs}[1]{{\bf Supplementary Fig.~#1}}
\newcommand{\fga}[1]{\textcolor{linkcolor}{Fig.~\ref{#1}}a}
\newcommand{\fgb}[1]{\textcolor{linkcolor}{Fig.~\ref{#1}}b}

\newcommand{\pana}{({\bf a})\xspace}
\newcommand{\panb}{({\bf b})\xspace}
\newcommand{\panc}{({\bf c})\xspace}

\newcommand{\red}{\textcolor{red}	} 
\newcommand{\blue}{\textcolor{blue}	}

\definecolor{citecolor}{rgb}{0.071, 0.36, 0.67}   
\definecolor{linkcolor}{rgb}{0.071, 0.4, 0.67}  
\hypersetup{
colorlinks=true, 
citecolor=citecolor,
linkcolor=linkcolor,
urlcolor=linkcolor
}

\newcommand{\name}{HI\emph{quant}\xspace	}

\graphicspath{ 
				{Figs/}   
}

\let\citep=\autocite
\let\citet=\autocite
\date{}
\begin{document}

\begin{spacing}{1.8}
\noindent {\LARGE \bf 
Quantifying homologous proteins and proteoforms 
}

\end{spacing}
\vspace{10mm}

\noindent\ann{\small
Dmitry Malioutov,$^{1}$
Tianchi Chen,$^{2}$
Jacob Jaffe,$^{3}$
Edoardo Airoldi,$^{4}$ 
Steven Carr,$^{3}$ \\
Bogdan Budnik,$^{5}$ \& 
Nikolai Slavov,$^{2,}$\textsuperscript{\Letter }
} \\

\noindent 
\normalsize{$^{1}$T.~J. Watson IBM Research Center, 1101 Kitchawan Road, Yorktown Heights, NY 10598, USA} \\
\normalsize{$^{2}$Department of Bioengineering, Northeastern University, Boston, MA 02115, USA}\\ 
\normalsize{$^{3}$Proteomics Platform, Broad Institute of MIT and Harvard, Cambridge, MA 02142, USA}\\
\normalsize{$^{4}$Department of Statistics, Harvard University, Cambridge, MA 02138, USA}\\
\normalsize{$^{5}$FAS Center for Systems Biology, Harvard University, Cambridge, MA 02138, USA}\\ 
%

{{\Letter}  Correspondence should be addressed to:                  
\href{mailto:nslavov@alum.mit.edu}{\an{\small nslavov@alum.mit.edu}} }

\thispagestyle{empty}




\begin{spacing}{1.34}
\newpage
{\bf
Many proteoforms -- arising from alternative splicing, post-translational modifications (PTMs), or paralogous genes -- have distinct biological functions, such as histone PTM proteoforms. However, their quantification by existing bottom-up mass--spectrometry (MS) methods is undermined by peptide-specific biases. To avoid these biases, we developed and implemented a first-principles model (\name) for quantifying proteoform stoichiometries. We characterized when MS data allow inferring proteoform stoichiometries by \name, derived an algorithm for optimal inference, and demonstrated experimentally high accuracy in quantifying fractional PTM occupancy without using external standards, even in the challenging case of the histone modification code.  
A \name server is implemented at: \href{https://web.northeastern.edu/slavov/2014_HIquant/}{https://web.northeastern.edu/slavov/2014\_HIquant/}        
}
\vspace{1cm}

\def \model {Model for inferring stoichiometries among proteoforms and paralogous proteins independently from peptide-specific biases. 
\pana 
One shared ($x_2$) and three unique ($x_1$, $x_3$ and $x_4$) peptides of H3 proteoforms illustrate a very simple case of \name.
\name models the peptide levels measured across conditions ($\vec x$) as a supposition of the protein levels ($\vec p$), scaled by unknown peptide--specific biases/nuisances ($z$). These coupled equations can be written in a matrix form whose solution infers the methylation stoichiometry  independently from the nuisances ($z$).
\panb The general form of the model for K proteoforms (or homologous proteins) with M peptides quantified across N conditions can be formulated and solved. In many, albeit not all, cases an optimal and unique solution can be found, even in the absence of unique peptides; see \figs{1} and \supp.}

\def \flowchart  {Work flow of \name and quality control (QC). 
\pana  \name takes as input peptide levels quantified across conditions (without restrictions on the quantification method) and estimates: $(i)$ The inference error for the dataset as described in (b); (ii) Quantifies across conditions proteoforms and homologous protein that cannot be quantified distinctly only from unique peptides;  (iii) Quantifies the stoichiometry across proteoforms and homologous protein.    
\panb The subset of homologous proteins and proteoforms that can be quantified by either unique or shared peptides are used to evaluate the accuracy of \name, i.e., quality control. Their relative protein levels inferred \emph{only} from shared peptides correlate strongly to the corresponding levels estimated \emph{only} from unique peptides.  
\panc Distribution of correlations (as in panel b) between protein levels inferred only from unique peptides or only from shared peptides. }

\def \ups  {\name accurately quantifies ratios across alkylated proteoforms of a spiked-in standard.
\pana  Schematic diagram of a validation experiment. We prepared a gold standard of proteoforms from the dynamic universal proteomics standard (UPS2) whose cysteines were covalently modified either with iodoacetamide or with vinylpyridine. Upon digestion, these modified UPS proteins generate many shared peptides (peptides not containing cysteine) and a few unique peptides (peptides containing cysteine). The modified UPS2 proteins were mixed with one another at known ratios ($n$), mixed with yeast lysate, digested and quantified by MS. The proteoform ratios that \name inferred from the MS data ($\hat n$) were compared to the mixing ratios.
\panb  The ratios \emph{across} the alkylated isoforms of UPS2 inferred by \name ($\hat n$, y-axis)  accurately reflect the mixing ratios ($n$, x-axis). 
\panc Comparison of the error in proteoform ratios inferred by \name and ratios inferred from the precursor ion areas and the reporter ion (RI) ratios.}

\def \h3  {\name accurately infers stoichiometries and confidence intervals across PTM site occupancies of histone 3. 
\pana  Histone 3 peptides were quantified by SRM across 7 perturbations, and the fractional site occupancies for K4 methylation estimated by two methods: Estimates inferred by \name without using external standards are plotted against the corresponding estimates based on MasterMix external standards with known concentrations \citep{creech2015building}. Each marker shape corresponds to the PTM site(s) shown in the legend; methylation is denoted with ``me'' and acetylation with ``ac'' followed by the number of methyl/acetyl groups.   
\panb The validation method from (a) was extended to another set of more complex fractional site occupancies on K9 methylation and K14 acetylation. 
}

\def \RPLphospho {Model for inferring stoichiometries among proteoforms and paralogous proteins independently from peptide-specific biases. 
\pana One shared ($x_1$) and two unique ($x_2$ and $x_3$) peptides from the two paralogs of ribosomal proteins L6 illustrate the simplest case of \name.  \name models the peptide levels measured across two conditions ($\vec x$) as a supposition of the protein levels ($\vec p$), scaled by unknown peptide--specific nuisances ($z$). These coupled equations can be written in a matrix form whose solution infers the $p_1 / p_2$ stoichiometry  independently from the nuisances ($z$). 
\panb The shared and unique peptides of proteoforms (as illustrated by PDHA1 phospho-proteoforms)  can be modeled as in panel (a); 
\panc The matrix system from (a) generalizes to K proteoforms (and homologous proteins) with M peptides quantified across N conditions. In many, albeit not all, cases an optimal and unique solution can be found, even in the absence of unique peptides. See \supp for details.}

Alternative mRNAs splicing and post-translational modifications (PTMs) produce multiple protein isoforms per gene, termed proteoforms \citep{Kelleher2013proteoform}. 
Furthermore, protein isoforms can be produced by distinct but highly homologous open reading frames, i.e., paralogous genes. Despite having similar sequence, proteoforms and protein isoforms often have distinct, even opposite biological functions \citep{soria2014functional}. For examples: (i) some Bcl-x isoforms promote apoptosis while other Bcl-x isoforms inhibit apoptosis \citep{schwerk2005regulation}; (ii) the methylation of histone 3 can cause either transcriptional activation (lysine 4) or repression (lysine 9) depending on the modified lysine \citep{berger2007complex}; and (iii) pyruvate kinase isoforms have different metabolic regulation, activities, and roles in aerobic glycolysis \citep{tanaka1967crystallization, christofk2008m2, Slavov_exp}. 

Understanding such systems demands quantifying proteoform abundances. This demand has motivated the development of external standards that can afford high accuracy even for complex proteoforms \citep{creech2015building}. However, their wider use has been limited by expense and applied only to special cases that allow chemical modification of cell lysates, e.g., phosphorylation \citep{wu2011large} and acetylation \citep{weinert2014acetylation, baeza2014stoichiometry}.  In the absence of external standards, the quantification of complex proteoform stoichiometries remains very challenging  because the ratios between proteoform-specific peptides do not necessarily reflect the ratios between the corresponding proteoforms \citep{olsen2010quantitative}; precursor ion areas corresponding to the same phospho-site in the same sample can differ over 100-fold depending on the choice of protease \citep{Multiple-Protease-2015}. This discrepancy is because a measured peptide level (precursor ion area) depends not only on the abundance of the corresponding protein(s) but also on extraneous factors including protein digestion, peptide ionization efficiency, the presence of other co-eluting peptides, and chromatographic aberrations \citep{apex2006absolute, albert-heck2012protease, Multiple-Protease-2015}. These extraneous factors break the equivalence between the abundance of a peptide and its precursor ion area and thus make protein quantification much more challenging than DNA quantification by sequencing.  This problem is compounded when PTM peptides have been enriched, and thus their intensities scaled by unknown enrichment-dependent factors.  

To infer proteoform stoichiometry, we use a simple model that is illustrated in \fg{schemes}a with proteoforms of histone H3 and in \figs{1} with paralogous ribosomal proteins and phospho-proteoforms of pyruvate dehydrogenase.
%
%
\name explicitly models peptide levels measured across conditions as a superposition of the levels of the proteins from which the peptides originate, \fga{schemes}. In this model, shared peptides serve as indispensable internal standards; they couple the equations for different peptides and thus make possible estimating stoichiometries between homologous proteins and proteoforms.  The simple example in \fga{schemes} generalizes to any number of proteins / proteoforms (M) and any number of conditions greater than 1 ($N>1$) as the system in \fgb{schemes} shows. \name solves this system and infers the protein levels (P) independently from the extraneous noise ($Z$; coming from protein-digestion, peptide-ionization differences, sample loss during enrichment, and even coisolation interference); $Z$ is also inferred as part of the solution and discarded. 
A related superposition model has been used before with peptides quantified at one condition \citep{Sarah2013Statistical}. However for a single condition, the model cannot quantify the proteins independently from the nuisance Z since all problems described by system 1 in  \fg{schemes} are under-determined, i.e., have infinite number of solutions (Proof 1; \supp). Thus, for a single condition, the model cannot take advantage of the robust corresponding-ion pairs, i.e., ratios between ions with the same chemical composition. In contrast, \name  infers ratios across proteins and their PTMs solely from the corresponding-ion ratios. This is possible because when $N>1$, the system in \fgb{schemes} often has a unique solution up to a single scaling constant, even when all peptides are shared, e.g., the problem defined by the design matrix in \figs{1}c. We characterize the conditions under which \name has a unique solution for the abundances of individual proteoforms and derive algorithms that use convex--optimization to find the optimal solution given the data; see \citet{Slavov_STLS} and \supp.


Our model (\fgb{schemes}) aims to make proteoform quantification insensitive to many systematic biases. For example, incomplete cleavage of a peptide, e.g., only 5\% of the peptide is released during enzyme digestion, is fully absorbed into the corresponding nuisance and does not affect inferred protein levels as long as the cleavage is 5\% for all conditions/samples. Analogously, if coisolation interference compresses the fold-changes of a peptide, the systematic component of the compression is fully absorbed by the nuisances. 
Unlike systematic biases, random noise in the data is not absorbed by the nuisances; it can degrade the quality of the inference. In order to assess the reliability of the inferred proteoform abundances, \name carefully evaluates the inference and assigns confidence levels. The evaluation uses inference features, such as fraction of explained variance, eigenvalue spectrum spacing and noise sensitivity; see \supp.

We sought to experimentally evaluate \name's ability to infer the proteoform stoichiometry in samples for which stoichiometry is accurately determined by other methods. The first method included creating and mixing proteoforms. The second method included quantifying histone H3 proteoforms relative to heavy peptide standards with predetermined abundances.

We aimed to create proteoform mixtures with known stoichiometries so that they can be used to assess the accuracy of  stoichiometries inferred by \name. 
To this end, the dynamic universal proteomics standard (UPS2) was digested, and the peptides split into two equal parts, A and B. In part A, cysteines were covalently modified with iodoacetamide, and in part B with vinylpyridine, \fga{ups}. We mixed part A and B in predefined ratios ($n$) and spiked each mixing ratio into an yeast sample. All samples were labeled with TMT, and the relative peptide levels quantified from the reporter ions at the MS2 level.

These alkylated UPS proteoforms have mostly shared peptides (peptides not containing cysteine) and a few unique peptides (peptides containing cysteine). \name modeled the relative levels of these peptides as shown in \fg{schemes} and solved the model to infer the stoichiometries of the alkylated proteoforms ($\hat n$), which should correspond to the mixing ratios.    
A comparison between the actual mixing ratios ($n$) with the inferred ratios  ($\hat n$) demonstrates a median error below $10\; \%$ for ratios inferred by \name and a substantially larger error for the ratios between precursor ion areas of unique peptides (\fg{ups}b,c); see \supp.


Next, we sought to evaluate the ability of \name to infer stoichiometries of more complex PTM proteoforms, those of histone H3. This system allows rigorous quantification of endogenous proteoform stoichiometries by a previously developed external standards (MasterMix) with known concentrations \citep{creech2015building}. For the test, we used peptides quantified by selective reaction monitoring (SRM)  across 7 perturbations. Fractional site occupancies were either estimated based on the external standards or inferred by \name  \emph{only} from the relative levels of the indigenous peptides, without using the MasterMix concentrations. The good agreement between these estimates (\fg{h3}) validates the ability of \name to infer fractional site occupancy even when the same site may be modified by different PTMs. The estimates from the external standards and from \name are very close but also show some systematic deviations. Those deviations may arise due to incomplete protein digestion that is hard to control for with peptide standards, measurement noise corrupting the solution inferred by \name or proteforms not explicitly included in the model. The abundances of some proteoforms with quantified peptides is over 1000 fold lower than the abundance of the main proteoforms. They and their corresponding peptides were omitted from the \name inference since their quantification requires unrealistically high accuracy of relative quantification; see \supp and Discussion.

The idea of using ratios between chemically identical ions is a cornerstone of quantitative proteomics \citep{blagoev2004temporal}. It has been used for decades in the context of relative quantification of proteins based on unique peptides \citep{altelaar2013benchmarking} and even applied to the special case of inferring phosphorylation cite occupancy  \citep{olsen2010quantitative}. Our work expands and generalizes this idea to all peptides, to  stoichiometries of complex proteoforms, and to unlimited number of conditions. Crucially, \name allows accurate, efficient, and numerically stable inference resulting in reliability estimates.      

\name  requires and depends upon accurate relative quantification. This limitation is largely and increasingly mitigated by technological developments allowing accurate estimates of corresponding-ion ratios. However, these technological developments on their own do not allow accurate estimates of PTM site occupancy \citep{Multiple-Protease-2015}. \name's dependence on the accuracy of relative quantification increases with increasing difference in the abundance of proteoforms. If the levels of two proteins differ by more than 3-6 orders of magnitude, this difference is likely better inferred from the precursor ion areas of the unique peptides. The associated noise (due to variability in protein digestion and ionization) is generally below 100 fold \citep{albert-heck2012protease} and thus smaller than the signal. \name's utility is particularly relevant when proteins and proteoforms have comparable abundances (within 10-100 fold difference) but distinct functions \citep{Slavov_ribo} and thus accurate quantification is essential for quantifying relatively small differences in abundance. Quantifying proteoforms is an exciting frontier essential for understanding post-transcriptional regulation \citep{floor2016tunable,Franks2016PTR_PLoS} and defining cell-types from single cell proteomes \citep{scopems2017}.

The general form of \name described in \fg{schemes}c indicates that \name is not limited to proteoforms, even broadly defined. Rather, \name can be applied to any set of proteins sharing a peptide. Here we emphasize the application to proteoforms because existing bottom-up methods are better suited for quantifying the stoichiometry between proteins with low homology that generate many unique peptides. For proteins with multiple unique peptides, some of the peptide-specific bias (from variation in protein-digestion and peptide-ionization efficiency) is likely to be averaged out and reduced. However, this bias is a more serious problem for proteoforms with only one or only a few unique peptides \citep{Multiple-Protease-2015}. For such proteoforms, \name can allow estimating stoichiometries accurately using only ratios between chemically identical ions, and provide a measure of the internal reliability of its own estimated abundances. \\

\noindent{ \bf Supplemental Information.}
Supplemental information includes Extended Experimental Procedures, Mathematical Proofs, and Supplemental Figures can be found in the \supp. 
The supplemental website for interactive data analysis can be found at:\\ \href{https://web.northeastern.edu/slavov/2014_HIquant/}{https://web.northeastern.edu/slavov/2014\_HIquant/}  \\

\noindent{ \bf Acknowledgments.}
We thank M.~Jovanovic, S.~MacNamara, Y.~Katz, and MA.~Blanco  for critical discussions and feedback. N.S.~started this work while a postdoc with Alexander van Oudenaarden and thanks him for generous support. Research was funded by a grants to N.S. from NIGMS of the NIH under Award Number DP2GM123497, a SPARC grant from the Broad Institute to N.S and S.C, and a NEU Tier 1 grant to N.S.


\printbibliography

\newcommand{\phos}[1]{\ensuremath{\overset{\red{\textcircled{p}}}{\mbox{\bf #1}}}}
\newcommand{\meth}[1]{\ensuremath{\;\:\mathclap{ \overset{ \red{\bf \underset{|}{CH_3}} } {{\bf #1 }}}}\;\:}
\newcommand{\methI}[1]{\ensuremath{\;\:\mathclap{ \overset{ \red{\bf \underset{|}{(CH_3)_2}} } {{\bf #1 }}}}\;\:}
\newcommand{\shp}{\graybox{GGCAK}}
\newcommand{\shpp}{\graybox{VVYLK}}
\newcommand{\HistN}{\graybox{YRPGTVALR}}
\sodef\ann{\fontsize{11}{14}\fontfamily{phv}\selectfont}{.05em}{0.5em plus1em}{.1em plus.1em minus.1em}

\newpage
\section*{Figure 1} 
\vspace{4mm}
\begin{figure}[h!]
 \noindent{\hspace{-3mm} {\large \bf \hvfont  a}
    \hspace{-2 mm} \an{H3K4 Methyl-proteoforms} \hspace{24 mm } \an{Model}  }\\  \vspace{1 mm}
    
	\begin{minipage}{0.45\linewidth}

	   $p_1 - \underbrace{TKQTAR}_{x_1}$          $\cdots$  $\underbrace{\HistN}_{x_2}$ \\[0.1em]
	   $p_2 - \underbrace{T\meth{K}QTAR}_{x_3}$   $\cdots$  $\underbrace{\HistN}_{x_2}$ \\[0.1em]
	   $p_3 - \underbrace{T\methI{K}QTAR}_{x_4}$  $\cdots$  $\underbrace{\HistN}_{x_2}$ \\ 	
	\end{minipage}
	\begin{minipage}{0.08\linewidth}

	\end{minipage}
	\begin{minipage}{0.45\linewidth}  
	  {\Large	
	   $$	   
	   \vec x_i  =  z_i \sum_{j \in \omega_i} \vec p_j	   
	   $$ }  
	\begin{flushleft}     
	{\Large $\vec x_i$ } -- { \small $i^{th}$ peptide levels across $N$ conditions}\\ 
	{\Large $z_i$ } -- { \small $i^{th}$ peptide-specific bias (nuisance)}\\
	{\Large $\omega_i$ } -- { \small Set of proteins containing the $i^{th}$ peptide}
	{\Large $\vec p_j$ } -- { \small $j^{th}$ protein levels across $N$ conditions}  
	\end{flushleft}	  		   
	\end{minipage}
	\vspace{-9 mm}
	{ \large
	\begin{align*}		 
		\underbrace{   \hspace{\textwidth} 
	     }_{\Downarrow}
	      	 \\
	    \hspace{-35mm} 	      
      	\begin{matrix}
		   \vec x_2  =  z_2 (\vec p_1+\vec p_2+\vec p_3)  	   
		   & \quad \quad  \vec x_1  =  z_1 \vec p_1 
		   & \quad \quad \vec x_3  =  z_3 \vec p_2  
		   & \quad  \quad \vec x_4  =  z_4 \vec p_3 \quad \quad \quad \\
		\end{matrix}
	\end{align*}
	}

	\vspace{6mm}
	\begin{minipage}{0.85\linewidth}
	{\hspace{-6mm} \large \bf \hvfont  b}\\
	\vspace{-11mm}
	{ 
	\begin{align*}\label{system1}
		 \underbrace{
		 \stackrel{\longleftarrow \; \mbox{\scriptsize conditions} \longrightarrow}{
		\begin{pmatrix}
				x_{11}  & \hdots  & x_{1N} \\
				\vdots  & \ddots  & \vdots \\
		 		x_{M1}  & \hdots  & x_{MN}
		 \end{pmatrix}
		 } 
		 }_{\underset{\mbox{\it \blue{Measured (data)}}}{\mbox{ \ann{Peptide levels} }}}
		 = 
		 \underbrace{
		 \stackrel{\longleftarrow \; \mbox{\scriptsize peptides} \longrightarrow}{
		\begin{pmatrix}
				z_1	  & \hdots  & 0 \\
				\vdots 				  & \ddots  & \vdots \\
		 		0  				  & \hdots  & z_M	
		 \end{pmatrix}
		 }
		 }_{\underset{\mbox{\it \blue{Unkown}}}{\mbox{ \ann{Peptide Biases} }}}	
		 \underbrace{
		 \stackrel{\longleftarrow \; \mbox{\scriptsize proteins} \longrightarrow}{
		 \begin{pmatrix}
				s_{11}  & \hdots & s_{1K} \\
				\vdots  & \ddots  & \vdots \\
		 		s_{M1}  & \hdots  & s_{MK} 
		 \end{pmatrix}}
		  }_{\underset{\mbox{\it \blue{Kown Proteoforms}}}{\mbox{ \ann{Design matrix} }}}
		 \underbrace{
		 \stackrel{\longleftarrow \; \mbox{\scriptsize conditions} \longrightarrow}{
		 \begin{pmatrix}
				p_{11}  & \hdots  & p_{1N} \\
				\vdots  & \ddots  & \vdots \\
		 		p_{K1}  & \hdots  & p_{KN}
		 \end{pmatrix} 
		 }
		 }_{\underset{\mbox{\it \blue{Results}}}{\mbox{ \ann{Protein levels} }}}
	\end{align*}
	}	
	\end{minipage}	
	\caption{ \model }
     \label{schemes}	
\end{figure}





\newpage
\section*{Figure 2}
\vspace{8mm}
\begin{figure}[h!]
	   \begin{overpic}
	   		[width = .99\textwidth]{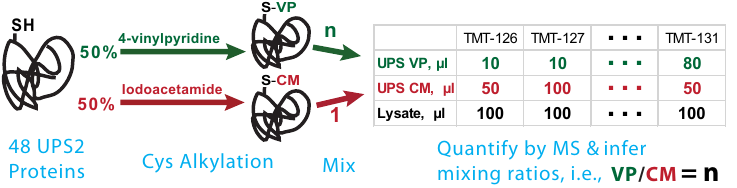} 
	              \put(-2,22){\large \bf \hvfont  a}
	   \end{overpic}
	\vspace{15mm}
	\\
	\inb{UPS_validation.pdf}{b}
	\hspace{0.08\textwidth}
	\inb{UPS_error4.pdf}{c} 
	\caption{ \ups }
	\label{ups}
\end{figure}

\newpage
\section*{Figure 3}
\vspace{8mm}
\begin{figure}[h!]
	\inb{H3K4.pdf}{a}
	\hspace{0.08\textwidth}
	\inb{H3K9-K14.pdf}{b}
	\caption{ \h3 } 
	\label{h3}
\end{figure}

\newpage       \setcounter{figure}{0}      \renewcommand{\thefigure}{ S\arabic{figure}}

\section*{Supplemental Figure 1}
\vspace{4mm}
\begin{figure}[h!]
	\begin{minipage}{0.45\linewidth}
	\noindent{\hspace{-3mm} {\large \bf \hvfont  a}
    \hspace{4 mm} \an{RP L6 Paralogs} }\\  \vspace{1 mm}
	   
	   $p_1 - \overbrace{\shpp}^{x_1}$  $\cdots$  $\overbrace{PH{\bf M}LK}^{x_2}$ \\[0.3em]
	   $p_2 - \underbrace{\shpp}_{x_1}$ $\cdots$  $\underbrace{PH{\bf L}LK}_{x_3}$ \\  \vspace{-6 mm} 

      	\begin{align*}
      	\underbrace{
      	\begin{matrix}
		& \vec x_1  =  z_1 (\vec p_1+\vec p_2)   &  \vec x_2  =  z_2 \vec p_1 
		& \vec x_3  =  z_3 \vec p_2  & \\
		\end{matrix}
	     }_{\big \Downarrow}
		\end{align*}
	 \vspace{-4 mm}  
	 {\small 
	 \begin{align*}	
	 \underbrace{
	  \begin{pmatrix}
			x_{11}  & x_{12}   \\
			x_{21}  & x_{22}   \\
	 		x_{31}  & x_{32}  \\
	 \end{pmatrix}
	 }_{\mbox{Peptide levels}}
	 = 	
	 	\underbrace{
	\begin{pmatrix}
			z_1	  & 0    & 0    \\
			0	  & z_2  & 0   \\
	 		0 	  & 0    & z_3  \\	
	 \end{pmatrix}
	 }_{\mbox{Nuisance}}
	 \underbrace{
	 \begin{pmatrix} 
			1  & 1 \\
			1  & 0 \\ 
	 		0  & 1 \\
	 \end{pmatrix}
	  }_{\mbox{\normalsize S}}
	 \underbrace{
	 \begin{pmatrix}
			p_{11}  & p_{12}   \\
	 		p_{21}  & p_{22} 
	 \end{pmatrix} 
	 }_{\mbox{Protein levels}}
	\end{align*}
	}	
	\end{minipage}
	\begin{minipage}{0.03\linewidth}

	\end{minipage}
	\begin{minipage}{0.45\linewidth}
	   \noindent{\hspace{-3mm} {\large \bf \hvfont  b}
    \hspace{4 mm} \an{Phospho-proteoforms} }\\  \vspace{1 mm} 
    \begin{flushright}
	   $p_1 - \overbrace{\shp}^{x_1}$  $\cdots$  $\overbrace{YGMGTSVER}^{x_2}$ \\[0.3em]
	   $p_2 - \underbrace{\shp}_{x_1}$ $\cdots$  $\underbrace{YGMG\phos{T}SVER}_{x_3}$ \\[0.3em]
	   $p_3 - \overbrace{\shp}^{x_1}$  $\cdots$  $\overbrace{YGMGT\phos{S}VER}^{x_4}$ \\[0.3em]
	   $p_4 - \underbrace{\shp}_{x_1}$  $\cdots$  $\underbrace{YGMG\phos{T}\phos{S}VER}_{x_5}$ \\
	\end{flushright}
	   \vspace{8 mm}
	\end{minipage}

    \vspace{6mm}
	\begin{minipage}{0.85\linewidth}
	{\hspace{-6mm} \large \bf \hvfont  c}\\
	\vspace{-11mm}
	{\scriptsize \small 
	\begin{align*}
		 \underbrace{
		 \stackrel{\longleftarrow \; \mbox{conditions} \longrightarrow}{
		\begin{pmatrix}
				x_{11}  & \hdots  & x_{1N} \\
				\vdots  & \ddots  & \vdots \\
		 		x_{M1}  & \hdots  & x_{MN}
		 \end{pmatrix}
		 } 
		 }_{\mbox{Peptide levels}}
		 = 
		 \underbrace{
		\begin{pmatrix}
				z_1	  & \hdots  & 0 \\
				\vdots 				  & \ddots  & \vdots \\
		 		0  				  & \hdots  & z_M	
		 \end{pmatrix}
		 }_{\mbox{Nuisance}}	
		 \underbrace{
		 \stackrel{\longleftarrow \; \mbox{proteins} \longrightarrow}{
		 \begin{pmatrix}
				s_{11}  & \hdots & s_{1K} \\
				\vdots  & \ddots  & \vdots \\
		 		s_{M1}  & \hdots  & s_{MK}
		 \end{pmatrix}}
		  }_{\mbox{Design matrix (S)}}
		 \underbrace{
		 \stackrel{\longleftarrow \; \mbox{conditions} \longrightarrow}{
		 \begin{pmatrix}
				p_{11}  & \hdots  & p_{1N} \\
				\vdots  & \ddots  & \vdots \\
		 		p_{K1}  & \hdots  & p_{KN}
		 \end{pmatrix} 
		 }
		 }_{\mbox{Protein levels}}
	\end{align*}
	}	
	\end{minipage}	
	\caption{ \RPLphospho }
     \label{RPLphospho}	
\end{figure}

\end{spacing}
\end{document}